\documentclass[10pt]{article}
\usepackage{graphicx}

\begin{document}
\title{On atomic state purity operator, degree of state purity and concurrence in the JC and anti-JC models}
\author{Joseph Akeyo Omolo\\
Department of Physics, Maseno University, P.O. Private Bag, Maseno, Kenya\\
e-mail:~ ojakeyo04@yahoo.co.uk}
\date{10 April 2022}
\maketitle

\begin{abstract}
The state of an atom in a bipartite qubit, Jaynes-Cummings (JC) or anti-Jaynes-Cummings (aJC) interaction is described by a reduced density operator. The purity of the state has been measured by taking the trace of the square of the reduced density operator. In this article, we define the square of the reduced density operator as the state purity operator, composed of a completely pure state part and a completely mixed state part. The coefficient of the completely mixed state part is the mixed state measure, formally obtained as the determinant of the reduced density operator and it is therefore directly related to tangle, the square of concurrence of the bipartite system. Expressed in various equivalent forms, the mixed state measure provides all the characteristic elements of state purity or entanglement, such as eigenvalues of the reduced density operator, nonclassicality measures and a state purity complex amplitude. The argument of the state purity complex amplitude in polar form is the phase of the state purity measure, which defines the degree of purity of the state. We find that the degree of purity and concurrence are complementary quantifiers satisfying a complementarity relation. The general form of the mixed state measure provides an interpretation that concurrence is fully defined by the Bloch radius four-vector in a spacetime frame. Plots of the degree of purity, concurrence and spin excitation number at resonance reveal that the atomic state collapses rapidly to a momentary totally mixed or maximally entangled state over a very short time after the beginning of the interaction, then evolves gently to a pure disentangled state in the middle of the collapse region of the spin excitation number. In off-resonance dynamics with large values of the frequency detuning parameters, the degree of purity, concurrence and spin excitation number develop periodic evolution at the respective maximum ($1$) or minimum ($0$) values, simultaneously signifying that at very large detuning, the atom evolves to a perfectly pure disentangled state where the spin excitation number is effectively zero.

\end{abstract}

\section{Introduction}
Atomic state purity or entanglement dynamics in the Jaynes-Cummings (JC) model is an old, comprehensively researched topic. The original studies initiated by Gea-Banacloche [1-3] and comprehensively elaborated by P L Knight and collaborators [4-6] revealed the important underlying property that the atom-field state evolves to a pure disentangled state in the middle of the initial collapse region of the atomic state population inversion. Introducing the reduced density operator $\rho$ to describe the state of the atom subsystem, Gea-Banacloche used the atomic state purity measure $Tr\rho^2$, while P L Knight and collaborators, used the von Newmann entropy defined in terms of the eigenvalues of $\rho$ to study the dynamical evolution of the atomic state purity or entanglement. These were essentially qualitative studies demonstrating the nature of the dynamical evolution between the pure disentangled and the maximally entangled states of the atom. It was established that the atomic state evolved rapidly from the initial pure state to a momentary maximally entangled state followed by a gentle evolution back to a pure disentangled state in the middle of the region of collapse of the state population inversion. The approximate forms of the emergent pure disentangled states of the atom and field mode were determined at half the revival time of the population inversion.

Generalizations beginning with the classic works of Bennette, et al [7], Hill and Wootters [8 , 9] shifted attention from the qualitative features of state purity or entanglement dynamics to a practical interpretation of entanglement of a bipartite system as a quantifiable quantum resource. Entanglement of formation of a pure state bipartite system was introduced as a quantifier defined as the von Newmman entropy defined in terms of the eigenvalues of the reduced density operator of either of the two subsystems or the binary entropy function defined in terms of a quantifier identified as the concurrence of the bipartite system. Hill and Wootters [8 , 9] provided an exact formula for the entanglement of formation by introducing concurrence as the basic quantifier of the entanglement of pure or mixed states of bipartite systems. The Hill-Wootters definition of concurrence, based on a spin-flip symmetry operation, was further developed in general forms by Uhlmann [10] through an arbitrary state conjugation symmetry operation and Rungta, et al [11] through a universal state inversion symmetry operation, while Albeverio and Fei [12] derived a general form of concurrence based on invariant traces of moments of the reduced density operator of a subsystem. In generalizations to distributed entanglement in multipartite systems, Coffman, Kundu and Wootters [13] introduced tangle as the natural quantifier, being the square of concurrence for a pure state bipartite system. In [14], Wootters presented an excellent review of the definition of entanglement of formation and the derivation of exact formulas of concurrence based on the state conjugation symmetry operations introduced in [8-11]. Concurrence has taken a central position in studies of entanglement of multipartite systems and current research interest is now focussed on lower bounds and monogamy of concurrence of mixed state multipartite systems [15-19].

In the midst of the current intensive investigations into the detailed features of concurrence, a recent information-theoretic approach based on the Wigner-Yanase skew information yielded yet another quantifier of entanglement, identified as the atomic nonclassicality quantifier [20]. The atomic nonclassicality quantifier has been used to study the evolution of the atomic state in the JC model [21], thereby revealing the development of a pure disentangled state in the middle of the collapse region of the spin state population inversion discovered earlier in [1 , 4 , 6]. In the present article, we have established that this nonclassicality quantifier is one of a pair of quantifiers of entanglement which can be interpreted as a redefinition of the concurrence of the atomic state.

In this article, we present a simplified approach to an effective theoretical study of the characteristic features of atomic state purity or entanglement dynamics in the JC , aJC and two-qubit interaction models. The underlying principle is that in an interacting bipartite system, the dynamics of a subsystem is effectively described by the reduced density operator of the subsystem. Hence, in contrast to the well established general approaches based on the bipartite system state conjugation symmetry operations developed in [8-11], we apply an interpretation that the dynamical evolution of the state purity or entanglement of the atom as a subsystem is fully described by a state purity operator defined as the square of the reduced density operator of the atom. Denoting the reduced density operator by $\rho$, applying the standard property that $\rho$ is hermitian ($\rho^\dagger=\rho$) and positive semi-definite ($\rho\ge0$) provides the state purity operator as $\rho^2$. If the atom is in a normalized pure state, then the state purity operator satisfies the basic property $\rho^2=\rho$. In general, the state purity operator is governed by an operator inequality $\rho^2\le\rho$, where the equality applies to a pure state, while for a mixed state, $\rho^2 < \rho$. Indeed, using the reduced density operator, it is established below that the state purity operator of the atom in the JC and aJC models satisfies the expected mixed state inequality. The state purity operator $\rho^2$ is composed of a completely pure state part and a completely mixed state part. We interpret the coefficient of the completely mixed state part as the mixed state measure which quantifies the purity or entanglement of the atomic state. The mixed state measure directly defines the square of the concurrence of the bipartite system, while expressing it in various equivalent forms provides all the characteristic elements of state purity or entanglement, namely, the reduced density operator eigenvalues, nonclassicality quantifiers and the complex amplitude of the state purity measure. The argument of the complex amplitude in polar form is the phase of the state purity measure, which defines the degree of purity of the state of the atom. A complementarity relation unifying the degree of purity and concurrence as complementary quantifiers of state purity or entanglement easily emerges from the form of the mixed state measure.

We begin with a brief description of the JC and aJC dynamics in section $2$, where we derive the general form of the reduced density operator of the atom in each model. In section $3$, we derive the state purity operator and determine the mixed state measure, which provides all the characteristic elements of state purity or entanglement ; we derive a direct relation between the mixed state measure and concurrence, then derive the degree of purity of the atomic state and the complementarity relation between the degree of purity and concurrence. We provide plots of the degree of purity, concurrence and the atomic spin excitation number to demonstrate the dynamical evolution of the state of the atom over time in the JC and aJC models. The Conclusion is presented in section $4$. Eigenvectors and eigenvalues of the reduced density operator are calculated in the Appendix.

\section{Atomic state in the JC and aJC models}
The JC and aJC models are the rotating and antirotating components of the quantum Rabi model of a two-level atom interacting with a single mode of quantized electromagnetic field. The JC and aJC Hamiltonians $H_{JC}$ , $H_{aJC}$ expressed in terms of the respective conserved excitation number and qubit state transition operators ($\hat{N} ~,~ \hat{R}$) , (~$\hat{\overline{N}} ~,~ \hat{\overline{R}}$) take the form
$$H_{JC}=\hbar\omega\hat{N}+\hbar\hat{R} \quad ; \quad \hat{N}=\hat{a}^\dagger\hat{a}+s_+s_- \quad ; \quad
\hat{R}=\delta s_z+g(\hat{a}s_++\hat{a}^\dagger s_-) \quad ; \quad \delta=\omega_0-\omega \eqno(1)$$
$$H_{aJC}=\hbar\omega\hat{\overline{N}}+\hbar\hat{\overline{R}} \quad ; \quad \hat{\overline{N}}=\hat{a}\hat{a}^\dagger+s_-s_+ \quad ; \quad \hat{\overline{R}}=\overline{\delta}s_z+g(\hat{a}s_-+\hat{a}^\dagger s_+) \quad ; \quad \overline{\delta}=\omega_0+\omega \eqno(2)$$
where $s_z , s_- , s_+ , \sigma_x=s_-+s_+ , \omega_0$ and $\hat{a} , \hat{a}^\dagger , \omega$ are the respective atomic spin-$\frac{1}{2}$ and field mode state transition operators and angular frequencies in standard definition, while $\delta$ , $\overline{\delta}$ are the red-sideband and blue-sideband frequency detunings. Noting the commutation relations $[~\hat{N}~,~\hat{R}~]=0$ , $[~\hat{\overline{N}}~,~\hat{\overline{R}}~]=0$, the time evolution operators $U_{JC}(t)$ , $U_{aJC}(t)$ generated by the JC and aJC Hamiltonians $H_{JC}$ , $H_{aJC}$ are expressed in the appropriate factorized form
$$U_{JC}(t)=e^{-i\hat{R}t}e^{-i\omega\hat{N}t} \ ; \quad\quad U_{aJC}(t)=e^{-i\hat{\overline{R}}t}e^{-i\omega\hat{\overline{N}}t} \eqno(3)$$
Considering the atom initially in the ground state $\vert g\rangle$ and the field mode initially in a coherent state $\vert\alpha\rangle$, the composite atom-field initial state $\vert\psi_{g\alpha}\rangle$ and the general time evolving states in the JC , aJC models are obtained as
$$\vert\psi_{g\alpha}\rangle=\sum_{n=0}^\infty\frac{e^{-\frac{1}{2}\alpha^2}\alpha^n}{\sqrt{n!}}\vert gn\rangle $$ $$\vert\Psi_{g\alpha}(t)\rangle=\sum_{n=0}^\infty\frac{e^{-\frac{1}{2}\alpha^2}\alpha^n}{\sqrt{n!}}U_{JC}(t)\vert gn\rangle \ ; \quad\quad \vert~\overline\Psi_{g\alpha}(t)\rangle=\sum_{n=0}^\infty\frac{e^{-\frac{1}{2}\alpha^2}\alpha^n}{\sqrt{n!}}U_{aJC}(t)\vert gn\rangle \eqno(4)$$
after expressing $\vert\alpha\rangle$ as a superposition of the field mode number (Fock) states $\vert n\rangle$. In this work, we take the coherent state eigenvalue $\alpha$ constant and real ($\alpha^*=\alpha$).

The JC , aJC excitation number and qubit state transition operators ($\hat{N} ~,~ \hat{R}$) , (~$\hat{\overline{N}} ~,~ \hat{\overline{R}}$) acting on $\vert gn\rangle$ generate the respective basic qubit states $\{\vert gn\rangle~,~\vert\phi_{gn}\rangle\}$ ,
$\{\vert gn\rangle~,~\vert~\overline\phi_{gn}\rangle\}$ satisfying transition operations
$$\hat{N}\vert gn\rangle=n\vert gn\rangle \quad ; \quad \hat{R}\vert gn\rangle=R_n\vert\phi_{gn}\rangle \quad ; \quad \hat{R}\vert\phi_{gn}\rangle=R_n\vert gn\rangle \quad ; \quad \vert\phi_{gn}\rangle=-c_n\vert gn\rangle+s_n\vert en-1\rangle $$
$$R_n=g\sqrt{n+\frac{1}{4}\beta^2} \quad ; \quad\quad c_n=\frac{\delta}{2R_n} \quad ; \quad\quad s_n=\frac{g\sqrt{n}}{R_n} \ ; \quad\quad \beta=\frac{\delta}{g} \eqno(5)$$

$$\hat{\overline{N}}\vert gn\rangle=(n+1)\vert gn\rangle \quad ; \quad
\hat{\overline{R}}\vert gn\rangle=\overline{R}_{n+1}\vert~\overline\phi_{gn}\rangle \quad ; \quad
\hat{\overline{R}}\vert~\overline{\phi}_{gn}\rangle=\overline{R}_{n+1}\vert gn\rangle \quad ; \quad
\vert~\overline\phi_{gn}\rangle=-\overline{c}_{n+1}\vert gn\rangle+\overline{s}_{n+1}\vert en+1\rangle $$
$$\overline{R}_{n+1}=g\sqrt{n+1+\frac{1}{4}(\beta+2f)^2} \quad ; \quad\quad \overline{c}_{n+1}=\frac{\delta}{2\overline{R}_{n+1}} \quad ; \quad\quad \overline{s}_{n+1}=\frac{g\sqrt{n+1}}{\overline{R}_{n+1}} \ ; \quad\quad f=\frac{\omega}{g} \eqno(6)$$
In defining the respective Rabi frequencies $R_n$ , $\overline{R}_{n+1}$ of qubit oscillations in equations (5) , (6), we have conveniently redefined the red-sideband and blue-sideband frequency detunings $\delta$ , $\overline{\delta}=\delta+2\omega$ to introduce the dimensionless coupling parameters $\beta$ , $f$ as defined in equations (5) , (6), noting the general property $\omega\ne0$ , $g\ne0$ , $f>0$.

Substituting $U_{JC}(t)$ , $U_{aJC}(t)$ from equation (3) into equation (4) as appropriate and applying the qubit state transition operations from equations (5) , (6), the respective JC , aJC general time evolving state vectors $\vert\Psi_{g\alpha}(t)\rangle$ , $\vert~\overline\Psi_{g\alpha}(t)\rangle$ are obtained and reorganized (Schmidt decomposition) in the form
$$\vert\Psi_{g\alpha}(t)\rangle=\sum_{n=0}^\infty(~A_n(t)\xi_n(t)\vert g\rangle-i~A_{n+1}(t)\eta_{n+1}(t)\vert e\rangle~)\vert n\rangle \quad ; \quad \xi_{n}(t)=\cos(R_nt)+ic_n\sin(R_nt) $$
$$\eta_{n+1}(t)=s_{n+1}\sin(R_{n+1}t) \quad ; \quad\quad A_n(t)=\sqrt{P_n}e^{-i\omega nt} \quad ; \quad\quad
A_{n+1}(t)=\sqrt{P_{n+1}}e^{-i\omega(n+1)t} \eqno(7)$$

$$\vert~\overline\Psi_{g\alpha}(t)\rangle=\sum_{n=0}^\infty(~{\overline{A}}_{n+1}(t)\overline{\xi}_{n+1}(t)\vert g\rangle -
i~{\overline{A}}_n(t)\overline{\eta}_n(t)\vert e\rangle~)\vert n\rangle \ ; \quad
\overline{\xi}_{n+1}(t)=\cos(~\overline{R}_{n+1}t)+i\overline{c}_{n+1}\sin(~\overline{R}_{n+1}t) $$
$$\overline{\eta}_n(t)=\overline{s}_n\sin(~\overline{R}_nt) \quad ; \quad\quad {\overline{A}}_{n+1}(t)=\sqrt{P_n}e^{-i\omega(n+1)t} \quad ; \quad\quad {\overline{A}}_n(t)=\sqrt{P_{n-1}}e^{-i\omega nt} \eqno(8)$$
The parameters and Rabi frequencies ($c_n ~,~ s_{n+1} ~,~ R_{n+j}$) , ($\overline{c}_{n+1} ~,~ \overline{s}_n ~,~ \overline{R}_{n+j}$), $j=0 , 1$, are defined according to equations (5) , (6), while the photon number state probabilities $P_{n+j}$ in the coherent state are defined by
$$P_{n+j}=\frac{e^{-\frac{1}{2}\alpha^2}\alpha^{n+j}}{\sqrt{(n+j)!}} \ , \quad j=0 , 1 , -1 \eqno(9)$$
Note that for completeness, the general solutions include the time evolving global phase factors $e^{-i\omega nt}$, $e^{-i\omega(n+1)t}$ which appear in the respective probability amplitudes $(A_n(t) ~,~ A_{n+1}(t)~)$ , $(~\overline{A}_n(t) ~,~ \overline{A}_{n+1}(t)~)$.

The reduced density operators $\rho_{ag\alpha}(t)$ , $\overline\rho_{ag\alpha}(t)$ of the atom in the JC , aJC dynamics are obtained by tracing out the field mode states ($Tr_f$) from the respective bipartite atom-field density operators $\rho_{g\alpha}(t)$ , $\overline\rho_{g\alpha}(t)$ as
$$\rho_{g\alpha}(t)=\vert\Psi_{g\alpha}(t)\rangle\langle\Psi_{g\alpha}(t)\vert \ ; \quad\quad \rho_{ag\alpha}(t)=Tr_f\rho_{g\alpha}(t) $$
$$\overline\rho_{ag\alpha}(t)=\vert~\overline\Psi_{g\alpha}(t)\rangle\langle~\overline\Psi_{g\alpha}(t)\vert \ ; \quad\quad \overline\rho_{ag\alpha}(t)=Tr_f\overline\rho_{g\alpha}(t) \eqno(10)$$
Substituting $\vert\Psi_{g\alpha}(t)\rangle$ , $\vert~\overline\Psi_{g\alpha}(t)\rangle$ from equations (7) , (8) and reorganizing using symmetrization relation $2(aA+bB)=(a+b)(A+B)+(a-b)(A-B)$, provides the reduced density operators in the general form
$$\rho=\frac{1}{2}(r_0\sigma_0+{\bf r}\cdot\vec\sigma) \ ; \quad\quad R=(r_0~,~r_1~,~r_2~,~r_3) \ ; \quad\quad \Sigma=(\sigma_0~,~\sigma_1~,~\sigma_2~,~\sigma_3) \ ; \quad \sigma_0=I \eqno(11)$$
where $R=(r_0~,~{\bf r})$ , $\Sigma=(\sigma_0~,~\vec\sigma)$ are the Bloch radius and Pauli spin matrix four-vectors, noting that $\sigma_0=I$ is the $2\times2$ identity matrix. In the reorganization of the reduced atomic state density operators in the above form, the Pauli matrices have been defined in terms of the basic atomic spin states $\vert g\rangle$ , $\vert e\rangle$ as usual
$$\sigma_0=\vert e\rangle\langle e\vert+\vert g\rangle\langle g\vert \ ; \quad \sigma_3=\vert e\rangle\langle e\vert-\vert g\rangle\langle g\vert \ ; \quad \sigma_1=\vert e\rangle\langle g\vert+\vert g\rangle\langle e\vert \ ; \quad
\sigma_2=-i(\vert e\rangle\langle g\vert-\vert g\rangle\langle e\vert) \eqno(12)$$
In the JC dynamics described by the state vector $\vert\Psi_{g\alpha}(t)\rangle$ in equation (7), the components of the Block radius four-vector have been obtained in the general form
$$r_0=\sum_{n=0}^\infty\vert A_{n+1}\vert^2\vert\eta_{n+1}\vert^2+\sum_{n=0}^\infty\vert A_n\vert^2\vert\xi_n\vert^2 \ ; \quad\quad
r_3=\sum_{n=0}^\infty\vert A_{n+1}\vert^2\vert\eta_{n+1}\vert^2-\sum_{n=0}^\infty\vert A_n\vert^2\vert\xi_n\vert^2$$
$$r_1=-i\sum_{n=0}^\infty(A_{n+1}A_n^*\eta_{n+1}\xi_n^*-A_{n+1}^*A_n\eta_{n+1}^*\xi_n) \ ; \quad\quad
r_2=\sum_{n=0}^\infty(A_{n+1}A_n^*\eta_{n+1}\xi_n^*+A_{n+1}^*A_n\eta_{n+1}^*\xi_n) \eqno(13)$$
In the aJC dynamics described by the state vector $\vert~\overline\Psi_{g\alpha}(t)\rangle$ in equation (8), the components of the Bloch radius four-vector have been obtained in the general form
$$r_0=\sum_{n=0}^\infty\vert~\overline{A}_n\vert^2\vert~\overline{\eta}_n\vert^2 +
\sum_{n=0}^\infty\vert~\overline{A}_{n+1}\vert^2\vert~\overline{\xi}_{n+1}\vert^2 \ ; \quad\quad
r_3=\sum_{n=0}^\infty\vert~\overline{A}_n\vert^2\vert~\overline{\eta}_n\vert^2 -
\sum_{n=0}^\infty\vert~\overline{A}_{n+1}\vert^2\vert~\overline{\xi}_{n+1}\vert^2$$
$$r_1=-i\sum_{n=0}^\infty(\overline{A}_n~\overline{A}_{n+1}^*~\overline{\eta}_n~\overline{\xi}_{n+1}^* -
\overline{A}_n^*~\overline{A}_{n+1}~\overline{\eta}_n^*~\overline{\xi}_{n+1}) \ ; \quad\quad
r_2=\sum_{n=0}^\infty(\overline{A}_n~\overline{A}_{n+1}^*~\overline{\eta}_n~\overline{\xi}_{n+1}^* +
\overline{A}_n^*~\overline{A}_{n+1}~\overline{\eta}_n^*~\overline{\xi}_{n+1}) \eqno(14)$$
We observe that, in general, the Bloch radius four-vector components $r_j$ , $j=0 , 1 , 2 , 3$ are obtained as the mean values of the corresponding Pauli matrices $\sigma_j$ with respect to the reduced density operators $\rho=\rho_{ag\alpha}(t) ~,~ \overline\rho_{ag\alpha}(t)$ of the atom in the respective JC , aJC general time evolving states (~$\vert\Psi(t)\rangle=\vert\Psi_{g\alpha}(t)\rangle , \vert~\overline\Psi_{g\alpha}(t)\rangle$~) according to
$$r_j=Tr\sigma_j\rho(t)=\langle\Psi(t)\sigma_j\vert\Psi(t)\rangle \ ; \ j=0 , 1 , 2 , 3 \ ; \quad\Rightarrow\quad Tr\sigma_0\rho=Tr\rho=\langle\Psi(t)\vert\Psi(t)\rangle=r_0 \eqno(15)$$

\section{Atomic state purity operator and mixed state measure}
As presented in section $2$ above, the JC and aJC models are pure state bipartite systems in which the atom and field mode subsystems are left in mixed states described by the respective reduced density operators obtained in general form in equation (11). In this section, we provide a simple consolidated derivation of the state purity operator to determine the mixed state measure and the associated  entanglement quantifiers of the atom in the JC and aJC models. The state purity operator is defined as the square of the reduced density operator of the atom.

Squaring $\rho$ in equation (11), applying standard algebraic identity
$({\bf a}\cdot\vec\sigma)({\bf b}\cdot\vec\sigma)=({\bf a}\cdot{\bf b})\sigma_0+i\vec\sigma\cdot({\bf a}\times{\bf b})$ giving
$({\bf r}\cdot\vec\sigma)({\bf r}\cdot\vec\sigma)=r^2\sigma_0$ and eliminating
${\bf r}\cdot\vec\sigma$ from the result using equation (11), we obtain the atomic state purity operator $\rho^2$ in the form
$$\rho^2=r_0\rho-{\cal M}\sigma_0 \ ; \quad\quad {\cal M}=\frac{1}{4}(r_0^2-r^2) \ ; \quad\quad r=\vert{\bf r}\vert~=\sqrt{r_1^2+r_2^2+r_3^2} \eqno(16)$$
This is an important relation, which reveals that the atomic state purity operator $\rho^2$ is composed of a completely pure state part $\rho$ and a completely mixed state part ${\cal M}\sigma_0$, where we interpret the coefficient ${\cal M}$ as the mixed state measure.

Treating the completely mixed state part in equation (4) as the completely mixed state density operator $\rho_{mix}$ provides the definition of the mixed state measure ${\cal M}$ as in equation (16) through the trace in the form (noting $Tr\sigma_0=TrI=2$)
$$\rho_{mix}={\cal M}\sigma_0 \ ; \quad\quad {\cal M}=\frac{1}{2}Tr\rho_{mix} \eqno(17)$$
Expressing the reduced density operator $\rho$ in equation (11) in the matrix form provides the general definition of the mixed state measure as the determinant according to
$$\rho=\frac{1}{2}\left(\matrix{r_0+r_3 & r_1-ir_2\cr r_1+ir_2 & r_0-r_3 \cr}\right) \ ; \quad\quad {\cal M}={\rm det}\rho \eqno(18)$$
We obtain the state purity measure ${\cal P}$ defined as the trace of the purity operator $\rho^2$ in equation (16) in the form
$${\cal P}=Tr\rho^2~:\quad\quad {\cal P}=r_0^2-2{\cal M} \eqno(19)$$
where we have used $Tr\rho=r_0$ from equation (15). The mixed state measure ${\cal M}$ thus defines the state purity measure and characterizes the dynamical evolution of the atomic state purity. The value ${\cal M}=0$ at $\vert{\bf r}\vert=r_0$, giving ${\cal P}=r_0^2$, characterizes a pure state, while ${\cal M}=\frac{1}{4}r_0^2$ at $\vert{\bf r}\vert=0$, giving ${\cal P}=\frac{1}{2}r_0^2$, characterizes a maximally entangled state, where we note that the normalized state case $r_0=1$ gives standard values of the purity measure ($r_0=1 ~,~ {\cal P}=1 , \frac{1}{2}$) for pure or maximally entangled states.

In [1], Gea-Banacloche used the time evolving state purity measure ${\cal P}=Tr\rho^2$ to study the dynamical evolution of the atomic state between pure and maximally entangled states in the normalized ($r_0=1$) bipartite state of the JC model. Evolution of the purity measure revealed a pure disentangled state developing in the middle of the collapse region of the atomic spin state population inversion. This interesting dynamical feature is demonstrated through the evolution of the degree of purity, concurrence and spin state excitation number in both JC and aJC models in Fig.$1$ , Fig.$2$ below.

Having established that ${\cal M}$ essentially defines the state purity measure, we now determine its underlying physical meaning by demonstrating how it is related to all the other characteristic elements of state purity, including concurrence, tangle and the degree of purity, which are interpreted as quantifiers of the entanglement of formation of the bipartite system.

\subsection{Characteristic elements of state purity measure}
Factoring the bracket in the definition in equation (16), we express the mixed state measure in the form
$${\cal M}=\varepsilon_-\varepsilon_+ \ ; \quad\quad \varepsilon_-=\frac{1}{2}(r_0-\vert{\bf r}\vert) \ ; \quad\quad \varepsilon_+=\frac{1}{2}(r_0+\vert{\bf r}\vert) \eqno(20)$$
where $\varepsilon_-$ , $\varepsilon_+$ are the eigenvalues of the reduced density operator $\rho$ as determined in the Appendix.

Next, adding and subtracting $r_0^2$ inside the bracket in equation (16) and reorganizing as appropriate gives alternative equivalent relations
$${\cal M}=\frac{1}{4}r_0^2-\lambda_-\lambda_+ \ ; \quad\quad \lambda_-=\frac{1}{2}\left(r_0-\sqrt{r_0^2-r^2}\right) \ ; \quad\quad \lambda_+=\frac{1}{2}\left(r_0+\sqrt{r_0^2-r^2}\right) \eqno(21)$$
$${\cal M}=\frac{1}{2}(r_0^2-\vert\varepsilon\vert^2) \ ; \quad\quad \varepsilon=\frac{1}{\sqrt{2}}(r_0+i\vert{\bf r}\vert) \quad\Rightarrow\quad \varepsilon=\sqrt{{\cal P}}~e^{i\varphi} \ ; \quad \tan\varphi=\frac{\vert{\bf r}\vert}{r_0} \eqno(22)$$
where ${\cal P}=\frac{1}{2}(r_0^2+r^2)$ in equation (22) is the state purity measure obtained earlier in equation (19).

It follows from equations (20) , (21) , (22) that the mixed state measure ${\cal M}$ is a composite of the characteristic elements $\varepsilon_\mp$ , $\lambda_\mp$ , $\varepsilon$ ($\varphi=\tan^{-1}\frac{\vert{\bf r}\vert}{r_0}$), which have been variously identified as basic quantifiers of entanglement or nonclassicality of the atomic state in the normalized state ($r_0=1$) JC model [4-6 , 20 , 21].

We observe that, in [4-6], P L Knight and collaborators used the eigenvalues $\varepsilon_\mp$ obtained here in the general form in equation (20) to define the von Newmann entropy
$${\cal S}=-\varepsilon_-{\rm log}_2\varepsilon_--\varepsilon_+{\rm log}_2\varepsilon_+ \eqno(23)$$
to study the evolution of the atomic state purity in the JC model. The evolution of the von Newmann entropy revealed precisely the same dynamical features determined through the purity measure in the earlier studies of Gea-Banacloche [1-3]. Noting that the evolution of the von Newmann entropy essentially follows the form of evolution of $\varepsilon_-$ over the same time ranges, Phoenix and Knight [1] interpreted the eigenvalue $\varepsilon_-$ as the basic measure of atomic state purity in the JC model.

On the other hand, we identify the characteristic element $\lambda_-$ obtained here in general form in equation (21) as the atomic state nonclassicality quantifier $N(\rho)$ recently derived through the Wigner-Yanase skew information in [20] and applied in studying atomic nonclassicality in the JC model in [21]. In the normalized state $r_0=1$, the nonclassicality quantifier $N(\rho)$ was determined [20 , 21] in the form
$$r_0=1~:\quad\quad \lambda_-=\frac{1}{2}\left(1-\sqrt{1-r^2}\right) \ ; \quad\quad N(\rho)=\lambda_- \eqno(24)$$
Below, we establish that the characteristic elements $\lambda_\mp$, now identified with the nonclassicality quantifier $N(\rho)=\lambda_\mp$, may be interpreted as redefinitions of the concurrence of the bipartite system (equation (27)).

In contrast, the complex characteristic element $\varepsilon$ introduced here in equation (22), may be interpreted as the state purity complex amplitude and its argument $\varphi$ is the phase of the state purity measure. The state purity complex amplitude has never been determined in earlier studies. As defined in equation (22), the argument defined by $r_0\tan\varphi=\vert{\bf r}\vert$, interpreted here as the phase of the state purity measure, directly describes the evolution of the Bloch sphere, characterizing the distribution of the atomic states within or on the surface of the Bloch sphere. The evolution of the state purity phase satisfies the inequality $0\le\tan\varphi\le1$, giving $0\le\varphi\le\frac{1}{4}\pi$, which follows from the property that the length $\vert{\bf r}\vert$ of the $3$-component Bloch radius vector ${\bf r}=(r_1~,~r_2~,~r_3)$ takes a range of values $0\le\vert{\bf r}\vert\le r_0$. Note that, in general, the phase $\varphi$ can take a spectrum of values $\varphi=q\pi$ ($q=0 , 1 , 2 , 3 , ...$) satisfying $\tan\varphi=0$ which characterize a maximally entangled state or $\varphi=(2q+1)\frac{1}{4}\pi$ satisfying $\tan\varphi=1$ which characterize a pure state.

\subsection{Concurrence and tangle}
Taking the trace of the state purity operator $\rho^2$ in equation (16) once again and reorganizing as appropriate, we obtain an important relation
$$2{\cal M}=r_0^2-Tr\rho^2 \quad\Rightarrow\quad 4{\cal M}={\cal C}^2 \ ; \quad\quad {\cal C}=\sqrt{2(r_0^2-Tr\rho^2)} \eqno(25)$$
after introducing the concurrence ${\cal C}$ of the bipartite system [11 , 14]. Concurrence was introduced as a quantifier of the entanglement of formation of a pure state bipartite qubit system by Hill and Wootters in [8 , 9] and determined in the explicit form in equation (25) by Rungta, et al in [11], noting $r_0=\vert\Psi(t)\vert\Psi(t)\rangle$ takes value $r_0=1$ for normalized bipartite state.

Equations (18) , (25) provide the definition of concurrence in terms of the mixed state measure ${\cal M}$ and the determinant of the reduced density operatot $\rho$ in the form
$${\cal C}=2\sqrt{\cal{M}} \quad\Rightarrow\quad {\cal C}=2\sqrt{{\rm det}\rho} \eqno(26)$$
According to equations (16) , (26), the vanishing of ${\cal M}$ at $\vert{\bf r}\vert=r_0$ means ${\cal C}$ vanishes, leaving the system in a pure disentangled state, while the evolution of ${\cal M}$ to $\frac{1}{4}r_0^2$ at $\vert{\bf r}\vert=0$, yielding ${\cal C}=r_0^2$, leaves the system in a maximally entangled state.

Using $(r_0^2-r^2)=4{\cal M}={\cal C}^2$ in equation (21) reveals that the nonclassicality quantifiers $\lambda_\mp$ can be interpreted as redefinitions of concurrence according
$$\lambda_\mp=\frac{1}{2}(r_0\mp{\cal C}) \eqno(27)$$
which means that the nonclassicality quantifier $N(\rho)=\lambda_-$ ($r_0=1$) determined in [20 , 21] is essentially the concurrence of the pure state bipartite system.

In [13], Coffman, Kundu and Wootters (CKW) introduced tangle $\tau$, defined as the square of concurrence, as the appropriate quantifier of the entanglement of formation of a bipartite system. Equations (25) , (26) provide the tangle of the pure state bipartite system in the precise form obtained by Coffmann, Kundu and Wootters as
$$\tau={\cal C}^2~:\quad\quad \tau=4{\cal M} \quad\Rightarrow\quad \tau=4{\rm det}\rho \eqno(28)$$
It follows from equations (19) , (26) , (28) that, the state purity measure ${\cal P}$ and the established quantifiers of the entanglement of formation of a bipartite system such as concurrence ${\cal C}$, tangle $\tau={\cal C}^2$, are defined in terms of the mixed state measure ${\cal M}$, which according to equations (20) , (21) , (22), is a composite of the characteristic elements of the state purity measure. In addition, equation (22) reveals another property that the state purity measure ${\cal P}$, defined by ${\cal M}$ in equation (19), has a hidden phase $\varphi=\tan^{-1}\frac{\vert{\bf r}\vert}{r_0}$, which directly describes the evolution of the Bloch sphere through the states of the atom. The state purity phase may thus be a more fundamental quantifier of state purity or entanglement, as we demonstrate through the degree of purity derivable from ${\cal M}$.

\subsection{Degree of state purity and complementarity relation with concurrence}
Reorganizing the definition of the mixed state measure ${\cal M}$ in equation (16) and introducing the definition of $\varphi$ from equation (22), we obtain the relation
$${\cal M}=\frac{1}{4}r_0^2(1-\tan^2\varphi) \ ; \quad\quad 0\le\tan\varphi\le1 \quad\Rightarrow\quad 0\le\varphi\le\frac{1}{4}\pi \eqno(29)$$
which reveals that the mixed state measure ${\cal M}$ is determined by  phase $\varphi$ of the state purity measure. The relations in equations (26) , (28) mean that concurrence ${\cal C}$ and tangle $\tau={\cal C}^2$ are also determined by the phase $\varphi$.

Introducing $r_0=Tr\rho$ , ${\cal M}={\rm det}\rho$ from equation (18) in equation (29), we obtain the definition of the phase of the state purity measure in the form
$$\tan\varphi=\sqrt{1-\frac{4{\rm det}\rho}{(Tr\rho)^2}} \eqno(30)$$
which takes precisely the same form as the degree of polarization of light, where $\rho$ corresponds to the coherence matrix [22 , 23]. In a corresponding interpretation, we now identify $\tan\varphi$ as the degree of purity of the atomic state.

Substituting the concurrence ${\cal C}$ from equation (26) into equation (29) or (30) provides an important property that the degree of purity $\tan\varphi$ and concurrence are complementary quantifiers of the purity or entanglement of the atomic state satisfying the complementarity relation in the normalized state $Tr\rho=r_0=1$
$$\tan^2\varphi+{\cal C}^2=1 \eqno(31)$$
It follows easily from the range of values $0\le\tan\varphi\le1$ in equation (29) and the complementarity relation in equation (31) that both degree of purity and concurrence take complementary values in the range $\{ 0 ~,~ 1 \}$, e.g., $(\tan\varphi=1 ~,~ {\cal C}=0)$ , $(\tan\varphi=0 ~,~ {\cal C}=1)$. The complementarity property means that the degree of purity and concurrence evolve in reverse order, one increasing and the other decreasing with time as demonstrated in Fig.$1$ , Fig.$2$ in the JC and aJC dynamics below.

Using $1-{\cal C}^2=\tan^2\varphi$ from the complementarity relation in equation (31), the binary entropy function $H(\frac{1}{2}(1\pm\sqrt{1-{\cal C}^2}))$ introduced in [7 , 8 , 9] as the entanglement of formation of the bipartite system is obtained in terms of the degree of purity in the form
$$H=-\frac{1}{2}(1+\tan\varphi)~\log_2\frac{1}{2}(1+\tan\varphi)-\frac{1}{2}(1-\tan\varphi)~\log_2\frac{1}{2}(1-\tan\varphi) \eqno(32)$$
The derivation of the degree of purity, the complementarity relation with concurrence and the associated binary entropy function completes the specification of essentially all the characteristic quantifiers of the atomic state purity or entanglement provided within the definition of the mixed state measure ${\cal M}$. We may therefore interpret the mixed state measure as the natural universal quantifier of the atomic (subsystem) state purity or entanglement of formation of the bipartite atom-field system.

In the present article, we interpret the degree of purity $\tan\varphi$ and concurrence ${\cal C}$, both varying within the range of values $\{0~,~1\}$ governed by the complementarity relation in equation (31), as the basic complementary quantifiers of the purity or entanglement of the atomic state. We present the interesting characteristic features of the degree of purity and concurrence of the atomic state in the JC and aJC models in Fig.$1$--Fig.$4$ below.

\subsubsection{Evolution of the degree of purity and concurrence in the JC and aJC models}
According to the effective definition of the degree of purity $\tan\varphi$ in equation (22), exactly equal to the form in equation (30) using equation (18), we only need to determine the explicit forms of the components of the Bloch radius four-vector obtained in equations (13) , (14) to demonstrate the evolution of the degree of purity and concurrence in the JC , aJC models.

The JC Bloch radius four-vector components are obtained in explicit form by substituting the definitions given in equations (7) , (9) into equation (13), giving
$$r_0=1 \ ; \quad\quad r_3=\sum_{n=0}^\infty P_{n+1}s_{n+1}^2\sin^2(R_{n+1}t)-\sum_{n=0}^\infty P_n(\cos^2(R_nt)+c_n^2\sin^2(R_nt)) $$
$$r_1=-2\sum_{n=0}^\infty\sqrt{P_{n+1}P_n}s_{n+1}\sin(R_{n+1}t)(~\sin(\omega t)\cos(R_nt)+c_n\cos(\omega t)\sin(R_nt)~)$$
$$r_2=2\sum_{n=0}^\infty\sqrt{P_{n+1}P_n}s_{n+1}\sin(R_{n+1}t)(~\cos(\omega t)\cos(R_nt)-c_n\sin(\omega t)\sin(R_nt)~) \eqno(33)$$
The aJC Bloch radius four-vector components are obtained in explicit form by substituting the definitions given in equations (8) , (9) into equation (14), giving
$$r_0=1 \ ; \quad\quad r_3=\sum_{n=0}^\infty P_{n-1}\overline{s}_n^2\sin^2(~\overline{R}_nt) -
\sum_{n=0}^\infty P_n(\cos^2(~\overline{R}_{n+1}t)+\overline{c}_{n+1}^2\sin^2(~\overline{R}_{n+1}t)) $$
$$r_1=2\sum_{n=0}^\infty\sqrt{PnP_{n-1}}\overline{s}_n\sin(~\overline{R}_nt)(~\sin(\omega t)\cos(~\overline{R}_{n+1}t) -
\overline{c}_{n+1}\cos(\omega t)\sin(~\overline{R}_{n+1}t)~) $$
$$r_2=2\sum_{n=0}^\infty\sqrt{PnP_{n-1}}\overline{s}_n\sin(~\overline{R}_nt)(~\cos(\omega t)\cos(~\overline{R}_{n+1}t) +
\overline{c}_{n+1}\sin(\omega t)\sin(~\overline{R}_{n+1}t)~) \eqno(34)$$
where the evaluation $r_0=1$ in each case establishes the property that the respective JC , aJC state vectors $\vert\Psi_{g\alpha}(t)\rangle$ ,  $\vert~\overline\Psi_{g\alpha}(t)\rangle$ are normalized according to equation (15). The important feature which emerges here is that the coherence components $r_1$ , $r_2$ are modulated by the periodically time varying field mode frequency-dependent factors $\sin(\omega t)$ , $\cos(\omega t)$ from the respective free evolution global phase factors $e^{-i\omega n t}$ , $e^{-i\omega(n+1)t}$ of the general solutions in equations (7) , (8).

The normalization property $r_0=1$ now means that the degree of purity takes the simple form $\tan\varphi=\vert{\bf r}\vert=\sqrt{r_1^2+r_2^2+r_3^2}$, which directly describes the evolution of the Bloch sphere between the pure states at $\vert{\bf r}\vert=1$ and the maximally entangled (totally mixed) states at $\vert{\bf r}\vert=0$. Using $r_1$ , $r_2$ , $r_3$ from equations (33) , (34) provides the explicit form of the degree of purity,  which we substitute into the complementarity relation in equation (31) to determine the corresponding concurrence in the JC , aJC models. Choosing the coherent field mode eigenvalue $\alpha=7$ (mean photon number $\overline{n}=\alpha^2=49$) to compare with Gea-Banacloche's original studies [1], we have plotted the degree of purity $\tan\varphi$ (blue color), concurrence ${\cal C}$ (red color) and spin excitation number $\frac{1}{2}(1+r_3)$ (green color) at resonance $\beta=0$ and off-resonance choosing $\beta=60$ in Fig.$1$-Fig.$4$ in the JC , aJC models. In general, the plots over scaled time $\tau=gt$ reveal that the modulation of the coherence components $r_1$ , $r_2$ by the global phase factors $\cos{\omega t}$ , $\sin{\omega t}$ does not affect the dynamical evolution, in agreement with the standard assumption that only the interaction component of the Hamiltonian generates measurable dynamics.

Under resonance $\beta=0$, Fig.$1$ , Fig.$2$ in the JC , aJC models show that the degree of purity (blue color) falls rapidly from the initial pure state at $\tan\varphi=1$ to a momentary totally mixed (maximally entangled) state at $\tan\varphi\approx0$, then rises gently to a very closely pure disentangled state at $\tan\varphi\approx1$ in the middle of the collapse region of the spin excitation number (green color). On the other hand, the concurrence (red color) rises rapidly from the initial pure state at ${\cal C}=0$ to a momentary totally mixed (maximally entangled) state at ${\cal C}\approx1$, then falls gently to a very closely pure disentangled state at ${\cal C}\approx0$ in the middle of the collapse region, precisely at about half the collapse time of the spin excitation number where the corresponding maximum value of the degree of purity occurs. The evolution in opposite sense of the two complementary quantifiers of state purity or entanglement simultaneously reveal that, within a very short time after the interaction begins, the atomic state collapses rapidly to a momentary maximally entangled state then evolves gently to an approximately pure disentangled state after about half the collapse time of the spin excitation number. This interesting feature of the dynamical evolution of the atomic state, exhibited here by the degree of purity and concurrence, agrees precisely with the original results of Gea-Banacloche exhibited through the evolution of the state purity measure $Tr\rho^2$ [1-3] and the results of P L Knight and collaborators exhibited through the evolution of the eigenvalue $\varepsilon_-$ and von Newmann entropy ${\cal S}$ [4-6] in the JC model. Approximate forms of the disentangled pure state emerging at half the collapse time of the spin state population inversion were determined in these original studies.

\begin{figure}[ph]
\centering
\includegraphics[width=0.35\linewidth]{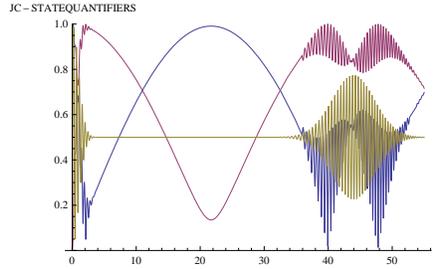}
\caption{Resonance evolution of atomic state degree of purity Tan$\varphi$ (BLUE), concurrence ${\cal C})$ (RED) and excitation number (GREEN) in the JC model at $\alpha=7~;~\beta=0~;~f=10^{-7}~{\rm over~ scaled~ time}~ \tau=gt$}
\label{Fig}
\end{figure}

\begin{figure}[ph]
\centering
\includegraphics[width=0.35\linewidth]{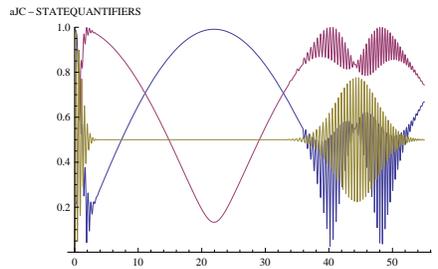}
\caption{Resonance evolution of atomic state degree of purity Tan$\varphi$ (BLUE), concurrence ${\cal C})$ (RED) and excitation number (GREEN) in the aJC model at $\alpha=7~;~\beta=0~;~f=10^{-7}~{\rm over~ scaled~ time}~ \tau=gt$}
\label{Fig}
\end{figure}

We have established that as the red-sideband detuning parameter $\beta$ is increased to large values, off-resonance dynamics in both JC and aJC models is characterized by periodic evolution of the degree of purity (blue color) with peaks (maximum values) at the upper value $\tan\varphi=1$, while the concurrence (red color) and the spin excitation number (green blobs on the zero-axis) each evolves periodically with deeps (minimum values) at the lower value ${\cal C}=0$ , $\frac{1}{2}(1+r_3)=0$ as shown in Fig.$3$ , Fig.$4$ for parameter values $\alpha=7$ , $\beta=60$ , $f=10^{-7}$. The periodicity is perfect such that the red deeps at the minimum values of the concurrence touch the corresponding green blobs of the spin excitation number on the zero-axis. The periodic evolution of the degree of purity, concurrence and the spin excitation number around their corresponding upper / lower values signifies the property that increasing the detuning parameter $\beta$ towards larger values progressively drives the dynamical evolution of the atom to a disentangled pure state. We find at very large values $\beta\ge175$ (with $\alpha^2=49$), the atom is in a perfectly pure state specified by  $\tan\varphi=1$ , ${\cal C}\approx0$ , $\frac{1}{2}(1+r_3)=0$. These features of the dynamical evolution to a pure disentangled state at large values of the detuning parameter $\beta$ agree precisely with the results of earlier studies of the JC model in [5]. The physical interpretation of this important dynamical feature follows from the definition of the dimensionless parameters $\beta=\frac{\delta}{g}$ , $f=\frac{\omega}{g}$, which specify the atom-field coupling regimes in the JC , aJC interactions. Increasing the parameters $\beta$ , $f$ to very large values essentially drives the system to very weak-coupling regimes where the atom-field interaction is too weak to generate state transitions, leaving the bipartite system in a disentangled pure state.

\begin{figure}[ph]
\centering
\includegraphics[width=0.35\linewidth]{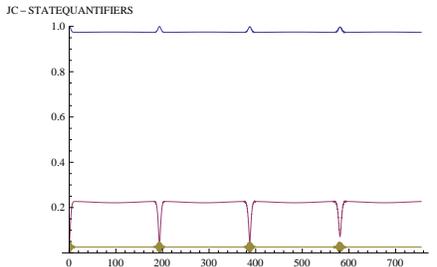}
\caption{Off-resonance evolution of atomic state degree of purity Tan$\varphi$ (BLUE), concurrence ${\cal C}$ (RED) and excitation number (GREEN) in the JC model at $\alpha=7~;~\beta=60~;~f=10^{-7}~{\rm over~ scaled~ time}~ \tau=gt$}
\label{Fig}
\end{figure}

\begin{figure}[ph]
\centering
\includegraphics[width=0.35\linewidth]{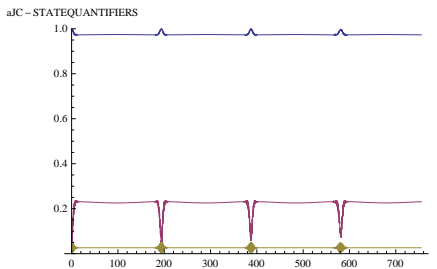}
\caption{Off-resonance evolution of atomic state degree of purity Tan$\varphi$ (BLUE), concurrence ${\cal C}$ (RED) and excitation number (GREEN) in the aJC model at $\alpha=7~;~\beta=60~;~f=10^{-7}~{\rm over~ scaled~ time}~ \tau=gt$}
\label{Fig}
\end{figure}
Even as we have described important features of the evolution of the atomic state in both JC and aJC models by varying only the dimensionless red-sideband detuning parameter $\beta$, which specifically arises in the JC interaction in equation (1), we must recall an underlying dynamical property that the aJC interaction is specified by the blue-sideband frequency detuning $\overline{\delta}=\omega_0+\omega$ as defined in equation (2). Hence, in specifying the general resonance and off-resonance conditions applicable to both JC and aJC dynamics, we have redefined the aJC blue-sideband frequency detuning $\overline{\delta}$ in terms of the JC red-sideband frequency detuning $\delta=\omega_0-\omega$ in equation (1) according to
$$\overline{\delta}=\delta+2\omega \ ; \quad \delta=\beta g \ ; \quad \omega=fg \quad\Rightarrow\quad \overline{\delta}=(\beta+2f)g \eqno(35)$$
where the dimensionless parameters $\beta$ , $f$ are defined in equations (5) , (6). It follows that the JC dynamics is characterized only by the parameter $\beta$, while the aJC dynamics is characterized by both $\beta$ and $f$. Hence, at resonance, the field-atom frequency $\omega=\omega_0$ is effectively eliminated from the Rabi frequency $R_{n+j}$ in the JC dynamics in equation (7), but remains as a residual detuning $\overline{\delta}_r=2fg$ , $f>0$ in the Rabi frequency $\overline{R}_{n+j}$ in the aJC dynamics in equation (8). In the plots in Fig.$1$-Fig.$4$, we have set $f=10^{-7}$ to minimize its contribution in the aJC relative to the JC dynamics. However, if we vary $f>0$ to large values, then even at resonance $\beta=0$, the atomic state evolution in the aJC can be driven to a disentangled pure state faster than in the JC, which will strictly follow the form of resonance evolution in Fig.$1$. In general, increasing both parameters $\beta$ , $f$ drives the atom to the pure disentangled state faster in the aJC compared to the JC evolution.

Apart from the property that the aJC process is faster than the corresponding JC process, it has emerged in the corresponding plots in Fig.$1$ , Fig.$2$, similarly, in Fig.$3$ , Fig.$4$, that the dynamical evolution of the atomic state takes the same form in the JC and aJC models. The degree of purity and the concurrence, which quantify state purity or entanglement, evolve in the same form in the two models. We observe that in the JC interaction the two-level atom couples to the \emph{rotating component}, while in the aJC interaction the atom couples to the \emph{counter-rotating component} of the same quantized electromagnetic field. Both interaction mechanisms have conserved excitation numbers and are characterized by transitions between well defined basic qubit states according to equations (1) , (2) , (5) , (6). The main difference between the two processes is that the JC interaction generates red-sideband transitions characterized by difference frequency $\delta=\omega_0-\omega$ equivalent to alternate emission or absorption by the atom or field mode, while the aJC interaction generates blue-sideband transitions characterized by sum frequency $\overline{\delta}=\omega_0+\omega$ equivalent to simultaneous emission or absorption by both atom and field mode. The dynamical evolution generated thus takes the same form. The similarity in the forms of dynamical evolution may also be associated with the property that the JC and aJC interaction Hamiltonians are duality conjugates, related by a unitary duality symmetry transformation.

\subsection{Concurrence and the Bloch radius four-vector}
Introducing the concurrence ${\cal C}=2\sqrt{{\cal M}}$ from equation (26) in equation (16) gives the important relation
$${\cal C}^2=(r_0^2-r^2) \quad\Rightarrow\quad {\cal C}^2=(r_0^2-{\bf r}\cdot{\bf r}) \ ; \quad\quad r=\vert{\bf r}\vert \eqno(36)$$
which we recognize as the four-vector mathematical relation for the invariant square of the length of the Bloch radius four-vector
$R=(r_0 ~,~ {\bf r})$ introduced in equation (11). This leads to the interpretation that the square of concurrence is equal to the square of the length of the Bloch radius four-vector defined in a four-dimensional spacetime frame. Introducing standard four-vector index $\mu=0 , 1 , 2 , 3$, we express the Bloch radius four-vector in the contravariant $R^\mu$ and covariant $R_\mu$ forms
$$R^\mu=(r_0 ~,~ {\bf r}) \ ; \quad\quad R_\mu=(r_0 ~,~ -{\bf r}) \ ; \quad\quad {\bf r}=(r_1 ~,~ r_2 ~,~ r_3) \ ; \quad \mu=0 , 1 , 2 , 3 \eqno(37)$$
The squared concurrence ${\cal C}^2$ in equation (36) is then obtained as the invariant squared length of the Bloch radius four-vector in the spacetime covariant mathematical form
$${\cal C}^2=R_\mu R^\mu=R^\mu R_\mu \quad\Rightarrow\quad {\cal C}=\vert R\vert=\sqrt{r_0^2-r^2} \eqno(38)$$
For normalized state $r_0=1$, we obtain
$$r_0=1~:\quad\quad R^\mu=(1~,~{\bf r}) \ ; \quad\quad R^2=R^\mu R_\mu=1-r^2 \ ; \quad\quad {\cal C}=\sqrt{1-r^2} \eqno(39)$$
It follows that concurrence is fully defined by the Bloch radius four-vector. Introducing the degree of purity in the form $r_0^2\tan^2\varphi=r^2$ from equations (22) , (30) into equation (38) provides the complementarity relation in equation (31) in the general form
$${\cal C}^2+r_0^2\tan^2\varphi=r_0^2 \eqno(40)$$
which reduces to the standard form in equation (31) for the normalized state $r_0=1$.

\section{Conclusion}
We have established that in a bipartite system of a two-level atom interacting with a single mode of quantized electromagnetic field in a JC , aJC model or with a spin-$\frac{1}{2}$ particle in a qubit-qubit model, the purity or entanglement of the state of the atom is described by a state purity operator obtained as the square ($\rho^2$) of the reduced density operator ($\rho$) of the atom. The state purity operator $\rho^2$ is composed of a completely pure state part and a completely mixed state part. The mixed state part is characterized by a mixed state measure, which can be expressed in various equivalent forms to provide all the elementary quantifiers of the purity or entanglement of the atomic state. The form of the mixed state measure automatically yields the complementarity relation which unifies the degree of purity and concurrence as complementary quantifiers of the state purity or entanglement. In addition, a simple interpretation that concurrence is fully defined by the Bloch radius four-vector is derivable from the general form of the mixed state measure, such that the square of concurrence is obtained in covariant form in a four-dimensional spacetime frame.

We have provided a qualitative picture by plotting the degree of purity, concurrence and the atomic spin excitation number to demonstrate the dynamical evolution of the atomic state over scaled time in the JC and aJC models. In resonance dynamics, the evolution of the degree of purity and concurrence between the corresponding extremal values in the range $\{0~,~1\}$ reveals that, the atomic state collapses rapidly to a momentary totally mixed or maximally entangled state within a very short time after the interaction begins, then evolves gently to a disentangled pure state in the middle of the collapse region of the spin excitation number, after which it falls gently to the revival point. In off-resonance dynamics, increasing the detuning parameters to very large values drives the degree of purity, concurrence and spin excitation number to periodic evolution along the respective extremal maximum or minimum values, signaling the evolution of the atom to a perfectly disentangled pure state where the degree of purity essentially takes the maximum value $1$, while the concurrence and spin excitation number each takes the corresponding minimum value $0$. By definition, increasing the detuning parameters progressively drives the atom-field system to weak-coupling regimes where beyond some critical large value, the interaction is too weak to generate transitions and the system remains in a perfectly pure state where the degree of purity is exactly $1$, while the concurrence and spin excitation number are exactly $0$.

\section{Acknowledgement}
I thank Maseno University for providing facilities and a conducive work environment during the preparation of the manuscript.

\section{Appendix : Eigenvectors and eigenvalues of the reduced density operator}
Rewriting the reduced density operator in equation (11) in the form ($\sigma_0=I$ , $\sigma_x=s_++s_-$ , $\sigma_y=-i(s_+-s_-)$)
$$\rho=\frac{1}{2}(r_0I+\hat{R}) \ ; \quad \hat{R}={\bf r}\cdot\vec\sigma=r_3\sigma_3+{\sqrt{r_1^2+r_2^2}}~(e^{-i\phi}s_++e^{i\phi}s_-) \ ; \quad\quad \phi=\tan^{-1}\frac{r_2}{r_1} \eqno(A1)$$
it is easily established that the state transition operator $\hat{R}$ acting on the atomic spin ground state $\vert g\rangle$ generates coupled qubit states
$\{\vert g\rangle ~,~ \vert\phi_g\rangle\}$ satisfying state transition operations
$$\hat{R}\vert g\rangle=\vert{\bf r}\vert\vert\phi_g\rangle \ ; \quad \hat{R}\vert\phi_g\rangle=\vert{\bf r}\vert\vert g\rangle \ ; \quad\quad \vert\phi_g\rangle=-\cos\theta~\vert g\rangle + \sin\theta~e^{-i\phi}~\vert e\rangle \ ; \quad \tan\theta=\frac{\sqrt{r_1^2+r_2^2}}{r_3} \eqno(A2)$$
Simple linear combination of the qubit states give the eigenvectors $\vert\psi_g^{\pm}\rangle$ of $\hat{R}$ and the density operator $\rho$,  satisfying eigenvalue equations
$$\vert\psi_g^\pm\rangle=\vert g\rangle \pm \vert\phi_g\rangle ~:\quad\quad \hat{R}\vert\psi_g^\pm\rangle=\pm\vert{\bf r}\vert\vert\psi_g^\pm\rangle \ ; \quad\quad \rho\vert\psi_g^\pm\rangle=\varepsilon_\pm\vert\psi_g^\pm\rangle \eqno(A3)$$
where the density operator eigenvalues $\varepsilon_\pm$ are exactly the characteristic elements of state purity or entanglement obtained in equation (20), which have been used to define the atomic state von Newmann entropy ${\cal S}$ in equation (23). It is also important to note that the Bloch radius $\pm\vert{\bf r}\vert$ are eigenvalues of the qubit state transition operator $\hat{R}$. Starting from the atomic spin excited state
$\vert e\rangle$ also provides the same eigenvalues. The eigenvectors $\vert\psi_g^\pm\rangle$ in equation (A3) are easily normalized as
$$\vert\psi_g^+\rangle=\cos\frac{1}{2}\theta\vert g\rangle +\sin\frac{1}{2}\theta e^{-i\phi}\vert e\rangle \ ; \quad\quad \vert\psi_g^-\rangle=\sin\frac{1}{2}\theta\vert g\rangle -\cos\frac{1}{2}\theta e^{-i\phi}\vert e\rangle \eqno(A4)$$
taking essentially the form, with phase $\phi$ defined in equation (A1), agreeing with the approximate eigenvectors and phase determined within the initial collapse period in the JC model in [4], particularly at $\cos(\frac{1}{2}\theta)=\sin(\frac{1}{2}\theta)=\frac{1}{\sqrt{2}}$. Notice that the phase $\phi$ is significantly different from the phase $\varphi$ of the state purity measure which we introduced in equation (22).

\end{document}